\documentclass[12pt,english]{article}
\usepackage[T1]{fontenc}
\usepackage[latin9]{inputenc}
\usepackage{amsmath}
\usepackage{amsthm}
\usepackage{amssymb}
\usepackage{graphicx}
\usepackage{setspace}
\makeatletter
\theoremstyle{plain}
\newtheorem{thm}{\protect\theoremname}
  \theoremstyle{remark}
  \newtheorem*{rem*}{\protect\remarkname}
  \theoremstyle{plain}
  \newtheorem{prop}[thm]{\protect\propositionname}
  \theoremstyle{plain}
  \newtheorem{cor}[thm]{\protect\corollaryname}
  \theoremstyle{remark}
  \newtheorem{claim}[thm]{\protect\claimname}
  \theoremstyle{plain}
  \newtheorem{lem}[thm]{\protect\lemmaname}

\makeatother

\usepackage{babel}
  \providecommand{\claimname}{Claim}
  \providecommand{\corollaryname}{Corollary}
  \providecommand{\lemmaname}{Lemma}
  \providecommand{\propositionname}{Proposition}
  \providecommand{\remarkname}{Remark}
\providecommand{\theoremname}{Theorem}

\begin{document}

\title{Quantum Leader Election}

\author{Maor Ganz\\
The Hebrew University, Jerusalem}

\date{October, 2016}
\maketitle
\begin{abstract}
A group of $n$ individuals $A_{1},\ldots A_{n}$ who do not trust
each other and are located far away from each other, want to select
a leader. This is the leader election problem, a natural extension
of the coin flipping problem to $n$ players. We want a protocol which
will guarantee that an honest player will have at least $\frac{1}{n}-\epsilon$
chance of winning\\
($\forall\epsilon>0$), regardless of what the other players do (weather
they are honest, cheating alone or in groups). It is known to be impossible
classically. This work gives a simple algorithm that does it, based
on the weak coin flipping protocol with arbitrarily small bias derived
by Mochon \cite{mochon-2007} in 2007, and recently published and
simplified in \cite{DoritAharonov2016}. A protocol with linear number
of coin flipping rounds is quite simple to achieve; We further provide
an improvement to logarithmic number of coin flipping rounds. This
is a much improved journal version of a preprint posted in 2009; The
first protocol with linear number of rounds, was achieved independently
also by \cite{dice} around the same time. 
\end{abstract}

\section{Introduction}

In this paper we present a quantum protocol for the Leader election
problem - a natural extension of coin flipping; in fact, our protocol
uses as a black box a quantum solution to the coin flipping protocol
(\cite{mochon-2007,DoritAharonov2016}). 

Thus, let us first review the coin flipping problem and what is known
about it. 

\subsection{The coin flipping problem}

A standard coin flipping is a game in which two parties, Alice and
Bob, wish to flip a coin from a distance. The two parties do not trust
each other, and would each like to win with probability of at least
$0.5$. A natural problem is to find good protocols - a protocol in
which a player could not cheat and force the outcome of the game to
his benefit.

There are two types of coin flipping - strong and weak. In strong
coin flipping, each party might want to bias the outcome to any result,
and the protocol has to protect against any such cheating. In weak
coin flipping each party has a favorite outcome, and so the protocol
has to protect only against cheating in that direction. 

We denote the winning probability of Alice to win a weak coin flipping
when both players are honest as $P_{A}$, and similarly $P_{B}$ for
Bob. The maximum winning probability of a cheating Alice (i.e. when
she acts according to her optimal strategy, while Bob is honest) is
denoted by $P_{A}^{*}$, and similarly $P_{B}^{*}$ for a cheating
Bob.\\
 Let $\epsilon=max(P_{A}^{*},\,P_{B}^{*})-\frac{1}{2}$ be the \textit{bias}
of the protocol. The bias actually tells us how good the protocol
is. The smaller the bias is, the better the protocol is.

It is well known that without computational assumptions, even weak
coin flipping is impossible to achieve in the classical world (see
\cite{Cleve87}. Note that impossibility of weak coin flipping implies
impossibility of strong coin flipping). That is, one of the players
can always win with probability $1$. In the quantum setting, the
problem is far more interesting.

\subsection{Quantum coin flipping}

Quantum strong coin flipping protocols with large but still non-trivial
biases were first discovered by \cite{Ahar} (with bias $\epsilon<0.4143$).
Kitaev then proved (see for example in \cite{ambainis-2003}) that
in strong coin flipping, every protocol must satisfy $P_{0}^{*}\cdot P_{1}^{*}\ge\frac{1}{2}$,
hence $\epsilon\geq\frac{\sqrt{2}-1}{2}$. This result raised the
question of whether weak coin flipping with arbitrarily small bias
is possible. Protocols were found with smaller and smaller biases
(\cite{ambainis-2003} showed strong CF with bias $\frac{1}{4}$,
\cite{Mochon2004.} showed weak CF with bias $0.192$), until Mochon
showed in his unpublished breakthrough paper \cite{mochon-2007} that
there are families of weak coin flipping protocols whose bias converges
to zero. This result was simplified in \cite{DoritAharonov2016}.
It is also known that even in the quantum world, a perfect protocol
(i.e. $\epsilon=0$) is not possible (\cite{D.Mayers1999a}).

\subsection{The leader election problem}

The leader election problem is the natural generalization of the weak
coin flipping, to $n$ players.\\
  The bias of the problem, is defined to be the minimal $\epsilon$
such that, every honest player has a winning probability of at least
$\frac{1}{n}-\epsilon$ (we do not have any limitation on the number
of cheating players). We will denote leader election with bias $\epsilon$
by $LE_{\epsilon}$.

As mentioned before, it is classically impossible to do a weak coin
flipping with a bias $<0.5$ without assumptions about the computation
power (\cite{Cleve87}). The same argument will show that it is also
impossible to solve the leader election problem in the classical setting,
in the sense that there will always be a player who can guarantee
getting elected. \\
Since leader election is not possible in the general sense, people
tried to use assumptions and conditions in order to make it possible.
For example, \cite{Feige-2000} presents a classical leader election
protocol (given that a player can flip a coin by herself) such that
given $\frac{(1+\delta)n}{2}$ honest players, the player whom the
protocol chooses is one of the honest players, with probability $\Omega(\delta^{1.65}).$
In addition there is a proof that every classical protocol has a success
(electing an honest leader) probability of $O(\delta^{1-\epsilon})$,
for every $\epsilon>0$. Note that there are limitation on the number
of cheaters.\\
Another variant is a protocol that chooses a processor randomly among
$n$ possibilities (In this case, there are no cheaters, the players
are anonymous and run the same protocol, and we want a protocol with
minimal running time / communication complexity, or one that works
without knowing the number of processors. See \cite{tani-2007} for
a quantum exact algorithm that solves this, but requires knowing a
bound on $n$). This type is sometimes called \textit{fair leader
election problem} (because there are no cheaters).

In this paper we investigate the version of Leader election with cheaters,
defined in the beginning of this subsection. We will refer to this
type simply as the \textit{leader election problem}. 

Until a short time before the first version of this result was posted,
there was no quantum result regarding this Leader election problem
as it was defined here. We note however that there were some results
on other types, such as \cite{ambainis-2003} which considered our
leader election problem, but allow penalty for cheaters that got caught
(which is obviously a weaker version of the problem), and also the
mentioned \cite{tani-2007} for the fair leader election problem.

\subsection{Our result}

If $n$ is a power of $2$, then a trivial solution exists, given
a good weak coin flip protocol. We can do a knock-out tournament (the
loser quits) of weak coin flipping, with $log(n)$ rounds. In each
round, all eligible players divide into pairs, and play weak coin
flip, where the loser gets eliminated from the tournament. The winner
of the tournament will be elected as the leader. 
\begin{thm}
\label{thm1:balanced tour}If $n=2^{k}$ for some $k\in\mathbb{N}$,
then for every $\epsilon>0$ there exists a leader election protocol
$LE_{\epsilon}$ , in which if all players are honest, then each has
a winning probability of $\frac{1}{n}$, otherwise any honest player
has a winning probability $>\frac{1}{n}-\epsilon$. With running time
of $O\left(N_{\epsilon}\log n\right)$, and $O\left(\log n\right)$
rounds of balanced weak coin flipping, where $N_{\epsilon}$ is the
number of rounds in a weak balanced coin flipping protocol $P_{\epsilon}$
of bias $\epsilon$.
\end{thm}
\begin{proof}
We will have full knock-out tournament of balanced weak coin flipping
$P_{\epsilon}$ between the players, and the winner of the tournament
will be declared as the leader.\\
It is obvious that if all players are honest, then each player has
a winning probability of $\frac{1}{n}=\left(\frac{1}{2}\right)^{k}$.\\
The only thing left to show is that each honest player has a winning
probability of at least $\frac{1}{n}-\epsilon$. \\
This is true because we have $\log n$ rounds, and in each round an
honest player has a winning probability of at least $\frac{1}{2}-\epsilon$,
hence in total the winning probability is at least
\[
\left(\frac{1}{2}-\epsilon\right)^{\log n}=\left(\frac{1-2\epsilon}{2}\right)^{k}\geq\frac{1-2\epsilon k}{2^{k}}=\frac{1}{n}-\epsilon\frac{2\log n}{n}\geq\frac{1}{n}-\epsilon
\]
where we used the Bernoulli inequality $\left(1+x\right)^{n}\geq1+nx$
which is true for any\\
 $x\geq-1\,,\,n\in\mathbb{N}$, and in our case, $\forall\ \epsilon\leq\frac{1}{2}$
(which is obviously satisfied).\\
We prove a generalization for that inequality in lemma \ref{lem:(a-x)}.
\end{proof}
A problem arises when $n$ is not a power of $2$, then this is not
possible, and putting in a dummy player involves some difficulties.
If the cheaters could control the dummy player, they would increase
their winning probability. This paper addresses this question.

Our first solution (which was also discovered in \cite{dice}, independently)
is to let $A_{1}$ play against $A_{2}$ and then the winner of that
will play $A_{3}$ and so on, as in a tournament, except we use unbalanced
weak coin flips. These are known to be possible using the balanced
weak coin flipping protocol (see in \cite{chailloux-2009}) but are
more expensive in terms of time.

As mentioned, in \cite{mochon-2007} Mochon showed the existence of
a weak coin flipping protocol with an arbitrarily small bias of at
most $\epsilon$. Let us denote this protocol by $P_{\epsilon}$ throughout
this paper. This protocol assumes that if both players are honest,
then each player has $\frac{1}{2}$ chance of winning (this is called
a \emph{balanced} coin flipping protocol). We denote by $N_{\epsilon}$
the running time of a balanced coin flipping protocol $P_{\epsilon}$
with bias $\epsilon$, which is the same as the number of rounds in
that protocol.\\
It is also possible to build an \emph{unbalanced} weak coin flipping
with an arbitrarily small bias $\epsilon$, in which if both players
are honest, then one honest player will have $q$ winning probability,
and the other player will have $1-q$ winning probability. If only
the first player is honest, his winning probability is at least $q-\epsilon$
(similarly $1-q-\epsilon$ for the second player, in case he is honest).
We will denote this protocol as $P_{q,\epsilon}$. In \cite{chailloux-2009}
it was shown that such protocols can be approximated using repetition
of $P_{\epsilon}$, with a total of $O\left(N_{\epsilon}\cdot\log\frac{1}{\epsilon}\right)$
rounds (See corollary \ref{cor:P_q,e} for more details).

The leader will be the winner of the final $\left(n-1\right)^{th}$
step. We arrive at the following theorem:
\begin{thm}
\label{thm2:exist}For every $\epsilon>0$, there exists a quantum
leader election protocol $LE_{\epsilon}$, in which any honest player
has a winning probability $\geq\frac{1}{n}-\epsilon$, with running
time of $O\left(n\cdot\log(\frac{n}{\epsilon})\cdot N_{\frac{\epsilon}{4n}}\right)$,
and $O\left(n\right)$ rounds of coin flipping.
\end{thm}
This theorem will be proven is section \ref{sec:Leader-election-protocol},
but as mentioned before, this simple solution is inefficient. It uses
a linear number of coin flipping rounds: $n-1$ of them.\\
Therefore we searched and found a better solution, that reduces the
number of coin flipping rounds, and yields the following theorem:
\begin{thm}
\label{thm3:efficient}For every $\epsilon>0$, there exists a leader
election protocol $LE_{\epsilon}$, in which any honest player has
a winning probability $\geq\frac{1}{n}-\epsilon$, with running time
of $O\left(N_{\frac{\epsilon}{4}}\log n\log\frac{1}{\epsilon}\right)$,
and $O\left(\log n\right)$ rounds of coin flipping.
\end{thm}
This theorem provides improvement of parameters with respect to theorem
\ref{thm2:exist}. The number of coin flipping rounds improves from
$O\left(n\right)$ to $O\left(\log n\right)$, and the number of total
unbalanced coin flipping from $O\left(n\right)$ to $O\left(\log n\right)$
as well. So the running time complexity (players can play in parallel
which does not increase the time complexity - we only count the longest
coin flipping in each round. See \ref{subsec:Definitions} for exact
definition) reduces from $\Theta\left(n\cdot\log(\frac{n}{\epsilon})\cdot N_{\frac{\epsilon}{4n}}\right)$
to $O\left(N_{\frac{\epsilon}{4}}\cdot\log n\cdot\log\frac{1}{\epsilon}\right)$
in the worst case, and in the best case, where $n=2^{k}$, our protocol
runs at $O\left(\log n\cdot N_{\epsilon}\right)$, achieving the optimal
complexity as that of Theorem \ref{thm1:balanced tour}. \\
The actual complexity improvement depends on $N_{\epsilon}$. As of
today, we only know a lower bound of $O\left(\log\log\frac{1}{\epsilon}\right)$
due to \cite{ambainis-2003}. In this scenario, assuming $\epsilon=\Theta\left(\frac{1}{n}\right)$
we get time complexity of $O\left(\log^{2}n\cdot\log\log n\right)$
instead of $O\left(n\cdot\log^{2}n\right)$, which is exponentially
better.\\
The only upper bound known is $\frac{1}{\epsilon}^{O\left(\frac{1}{\epsilon}\right)}$
due to \cite{DoritAharonov2016}. Even if we only use $N_{\epsilon}=O\left(2^{\frac{1}{\epsilon}}\right)$,
then assuming $\epsilon=\Theta\left(\frac{1}{n}\right)$ we get time
complexity of $O\left(2^{n}\cdot\log n\cdot\log\log n\right)$ instead
of $O\left(2^{n^{2}}\cdot n\log n\right)$, which is again exponentially
better.\\
However if $N_{\epsilon}$ is linear in $\frac{1}{\epsilon}$, then
assuming $\epsilon=\Theta\left(\frac{1}{n}\right)$ we get time complexity
$O\left(n\log^{2}n\right)$ instead of $O\left(n^{3}\log n\right)$.\\
Still in all cases theorem \ref{thm3:efficient} provides a significant
improvement in parameters. 

To prove theorem \ref{thm3:efficient}, we use a knock-out tournament
of weak coin flipping, in which the loser quits and the winner continues
to the next round. Since $n$ (the number of players) is not necessarily
a power of $2$, then one must adjust it.

Our protocol is fairly simple and uses $\log n$ rounds of unbalanced
weak protocols $P_{q_{i}^{\prime},\frac{\epsilon}{2}}$ as will be
defined later, in section \ref{sec:Improved-protocol}, where we will
prove Theorem \ref{thm3:efficient} (at most one at each round). This
limitation is important, because at the moment we only know how to
implement an unbalanced flip using a repetition of balanced coin flip,
which influences the total message complexity.

If it were possible to improve the complexity of unbalanced coin flip,
to that of a balanced coin flip (this is an open problem), then one
can improve the complexity of the suggested leader election protocol
to $O\left(N_{\epsilon}\log n\right)$. It is possible that this can
also be achieved by finding an appropriate families of time independent
point games (see \cite{mochon-2007,DoritAharonov2016}) to derive
more efficient protocols, but this has never been done before and
will not be done in this paper.

\subsection{Related work}

This work was first posted in 2009. This paper is the journal version
of that preprint, which is much improved. 

A related work was published \cite{dice} at the same time as the
preprint. They refer to the leader election problem as weak dice rolling,
and they use the same protocol as we did (independently) in \ref{first protocol},
proving theorem \ref{thm2:exist}. (However, they assume there exists
$P_{q,\epsilon}$ for every $q\in\left[0,1\right]$, which is not
known to be true, but only an approximation to such). As mentioned
before, our work improves that result significantly.\\
\cite{dice} also study the leader election problem in the strong
scenario, under the name of strong dice rolling. Namely they consider
the problem of $n\geq2$ remote parties, having to decide on a number
between $1$ and $N\geq3$, in which the parties want to avoid bias
in any direction. They generalize Kitaev's bound (see \cite{ambainis-2003})
to apply to $n$ parties $N$ sided strong dice-rolling. This was
done by noting that strong dice rolling can always be used to implement
strong imbalanced coin flipping. Note that this rules out Leader election
in the strong version for $n=N$.\\
\cite{dice} also extend the strong optimal coin flipping protocol
in \cite{chailloux-2009} and provide a family of strong dice rolling
protocols which matches this bound, for the case of the number of
parties being $n=2\cdot M$ and the number of outcomes to be $N=T^{M}$
for any $M,T\in\mathbb{N}$.

\subsection{Organization of the paper:}

Section \ref{sec:Leader-election-protocol} gives formal definitions
and proves Theorem \ref{thm2:exist}.\\
Section \ref{sec:Improved-protocol} proves Theorem \ref{thm3:efficient}.\\
Section \ref{sec:Open-questions} is open questions.\\
In the Appendix, we gives formal definition of weak coin flipping.

\section{\label{sec:Leader-election-protocol}Leader election protocol}

We start by some standard definitions.

\subsection{\label{subsec:Definitions}Definitions and requirements}

By a \textit{weak coin flipping protocol} we mean that Alice wins
if the outcome is $0$, and Bob wins if it is $1$.

A weak coin flipping protocol with $P_{A}=q,\ P_{B}=1-q$, bias $\epsilon$
will be denoted by $P_{q,\epsilon}$.

A \textit{leader election protocol} with $n$ parties $A_{1},\ldots,A_{n}$
has an outcome \\
$t\in\{1,\ldots,n\}$. We will denote by $P_{i}$ the probability
that the outcome is $t=i$.

We assume that each player has its own private space, untouchable
by other players, a message space $\mathcal{M}$ which is common to
all ($\mathcal{M}$ can include a space for the identification of
the sender and receiver).

We use the existence of a weak coin flipping protocol with bias at
most $\epsilon$ for every $\epsilon>0$ . This fact was proved in
\cite{mochon-2007} for $P_{A}=P_{B}=\frac{1}{2}$ and we will denote
it as $P_{\epsilon}$ and by $N_{\epsilon}$ the number of its rounds
(we define a round shortly). There is a proof that there is an approximation
to such a protocol for every $P_{A},\,P_{B}$ (s.t. $P_{A}+P_{B}=1$)
in \cite{chailloux-2009} by repetitions of $P_{\epsilon}$, with
$O\left(\log\frac{1}{\epsilon}\cdot N_{\epsilon}\right)$ rounds.
(It seems possible to generalize the weak coin flipping protocol {[}9,2{]}
directly to unbalanced coin flipping with any $P_{A}+P_{B}=1$, without
this $\log n$ increase in complexity, but this wasn't done yet.)
Recall that we denoted such unbalanced protocol with $P_{A}=q$ and
bias $\epsilon$ by $P_{q,\epsilon}$. We will denote the approximate
protocol by $P_{q^{\prime},\frac{\epsilon}{2}}$ See corollary \ref{cor:P_q,e}
for details. 
\begin{rem*}
A \emph{round} in a coin flipping protocol, consists of two steps:
First where Alice does something on her space $\mathcal{A}\otimes\mathcal{M}$
and sends Bob the message space. Second when Bob receives the message,
does something on his space $\mathcal{M}\otimes\mathcal{B}$, and
sends Alice a reply. Hence, the \emph{running time} of a coin flipping
protocol, is the number of its rounds (see section \ref{sec:Remarks}
for more details).\\
A \emph{coin flipping round} in our leader election protocol, is essentially
a coin flipping, performed in parallel, between the pairs of players
who are still eligible to be elected. \\
In our analogy to a knock-out tournament, a coin flipping round corresponds
to a round in that tournament.\\
The \emph{running time} of a coin flipping round is thus the running
time of a single coin flipping (which is the number of communication
rounds of the two players). Since different coin flipping in the same
round might take different times, we consider the worst coin flipping
time of each round.\\
Hence the running time of our leader election protocol, will be the
sum of the running times of its coin flipping rounds. It can be bounded
by the number of coin flipping rounds times the running time of the
worst coin flipping (in all the tournament).
\end{rem*}
\begin{prop}
\label{1. P_x_e_0 } \cite{chailloux-2009} Let $P_{\epsilon}$ be
a balanced weak coin flipping protocol with bias $\epsilon$ and $N_{\epsilon}$
rounds.\\
Then $\forall q\in[0,1]$ and $\forall k\in\mathbb{N}$, there exists
an unbalanced weak coin flipping protocol $P_{q^{\prime},\delta}$
with $k\cdot N_{\epsilon}$ rounds, such that $\mid q^{\prime}-q\mid\leq2^{-k}$
and $\delta=2\epsilon$.
\end{prop}
Assume we are interested in a protocol with $P_{A}=q\,,\ P_{B}=1-q$
for some $q$. \\
Let $k=1+\left\lceil \log\frac{1}{\epsilon}\right\rceil $. Then according
to the proposition w.r.t $P_{\frac{\epsilon}{4}}$, there exist $P_{q^{\prime},\delta}$
with $O\left(\log\frac{1}{\epsilon}\cdot N_{\frac{\epsilon}{4}}\right)$
rounds, such that $\mid q^{\prime}-q\mid\leq\frac{\epsilon}{2}$ and
$\delta=2\cdot\frac{\epsilon}{4}=\frac{\epsilon}{2}$. \\
Hence an honest first player has a winning probability of at least
\\
$q^{\prime}-\delta\geq q-\frac{\epsilon}{2}-\delta=q-2\delta$.\\
An honest second player has a winning probability of at least \\
$1-q^{\prime}-\delta\geq1-q-\frac{\epsilon}{2}-\delta=1-q-2\delta$.\\
Note that this protocol ensures that when both players are honest,
$P_{A}=q^{\prime}$, hence $\left|P_{A}-q\right|\leq\frac{\epsilon}{2}$.\\
Similarly $\left|P_{B}-\left(1-q\right)\right|\leq\frac{\epsilon}{2}\,,\ P_{A}+P_{B}=1$.\\
This proves:
\begin{cor}
\label{cor:P_q,e}For Every $q\neq\frac{1}{2}\,,\ $$\delta>0$ there
exists a weak coin flipping protocol $P_{q^{\prime},\delta}$ s.t.
$\left|q^{\prime}-q\right|\leq\delta$ with $O\left(N_{\frac{\delta}{2}}\cdot\log\frac{1}{\delta}\right)$
rounds.\\
When all players are honest, then $P_{A}=q^{\prime}$ is guaranteed
to satisfy \\
$\left|P_{A}-q\right|\leq\delta\,,\ P_{B}=1-q^{\prime}\,,\ \left|P_{B}-\left(1-q\right)\right|\leq\delta$.\\
The winning probability is at least $q^{\prime}-\delta\geq q-2\delta$
for an honest $A$ player, and $1-q-2\delta$ for an honest $B$ player.
\end{cor}
The proof of theorem \ref{thm2:exist} is basically a simple combinatorial
manipulation of how to combine balanced coin flips to achieve the
correct probability of winning. Corollary \ref{cor:P_q,e} enables
to calculate the complexity due to using balanced coin flips.

There is one delicate point: we have to make sure that the cheaters
can not increase their winning probability in a specific coin flip,
by losing previous coin flip (say by creating entanglements). This
will be discussed in subsection \ref{subsec:Group-of-cheaters}.

\medskip{}

We will first present the simpler case of three parties, to show the
basic idea. The general case is a natural generalization of this,
and we will analyze it in details. Nevertheless the three party case
captures the basic idea of the problem in its simple version, namely
the proof of Theorem \ref{thm2:exist}.

\subsection{\label{subsec:Group-of-cheaters}Group of cheaters}

When we analyze coin flipping between two players, we assume one of
them is honest and analyze the scenario that the other player is cheating
and we then bound his winning probability. In multiparty protocol,
such as the leader election, another possibility might occur. A group
of cheating players $A_{i}\in C$ might try to increase their winning
probability as a team. \\
For example, maybe it is possible that $A_{2},A_{3},A_{4}$ are a
cheating team. They know that in the first round $A_{1}$ plays $A_{2}$,
and in the second round the winner of that encounter will play $A_{3}$
or $A_{4}$. Maybe they can create some cheating strategy that will
cause $A_{2},A_{3}$ to lose, but will increase significantly the
winning chances of $A_{4}$ in the second round?\\
This might sound far fetched, but as \cite{sat} show, it is a possible
scenario in parallel coin flipping; it is the analogy of quantum hedging
(see \cite{WM}).\\
However, this is not possible to do in a sequential setting as ours.
The point is that an honest player $A_{j}$ plays one coin flipping
round at a time, and the next coin flipping protocol he participates
in starts after the previous one ends. In this setting, the following
claim holds: 
\begin{claim}
if $A_{j}$ is honest, and is playing coin flipping protocols sequentially,
where the $i^{th}$ protocol is $P_{q_{i}^{\prime},\epsilon_{i}}$,
against whoever may be, then her chance of winning the $i^{th}$ round,
even conditioned on whatever happened in previous rounds with other
players, is at least $q_{i}^{\prime}-\epsilon_{i}$. 
\end{claim}
\begin{proof}
Since $A_{j}$ is honest, she starts her $i^{th}$ round with a clean
ancilla register. Assume in the worst case that all other players
conspire against her together. Suppose they can collude to make her
winning probability strictly smaller than $q_{i}^{\prime}-\epsilon_{i}$.
Then a cheater playing against $A_{j}$ a single coin flipping protocol
can also do this, by simulating what they have done in previous rounds
to prepare the initial state of the protocol, and then simulating
what they do in the current round. This is in contradiction to the
fact that no matter what a cheating player does, $A_{j}$ has at least
$q_{i}^{\prime}-\epsilon_{i}$ probability to win against her opponent
when playing a single round of a coin flipping protocol $P_{q_{i}^{\prime},\epsilon_{i}}$. 
\end{proof}
In our protocols, every round each eligible player plays a weak coin
flipping $P_{q,\epsilon}$. This protocol guarantees the honest player
a winning probability of $q-\epsilon$.

\subsection{\label{subsec:Three-parties}Three parties: A,B,C}

Alice, Bob and Charlie want to select a leader.

Let $\epsilon>0$. We will show a leader election protocol $LE_{\epsilon}$,
such that an honest party has a winning probability of at least $\frac{1}{3}-\epsilon$.

Let $\epsilon^{\prime}=\frac{\epsilon}{2}$.

\subsubsection{Protocol}
\begin{enumerate}
\item Alice plays Bob $P_{\epsilon^{\prime}}$ (balanced coin flip with
bias $\frac{\epsilon}{2}$).
\item The winner plays Charlie $P_{\frac{2}{3}^{\prime},\epsilon^{\prime}}$,
where by this we mean $P_{q^{\prime},\epsilon^{\prime}}$ such that
$\left|q^{\prime}-\frac{2}{3}\right|\leq\frac{\epsilon}{2}$ (the
existence of this protocol is given by corollary \ref{cor:P_q,e}). 
\item The winner of that flip is declared as the leader.
\end{enumerate}
\includegraphics[scale=0.75]{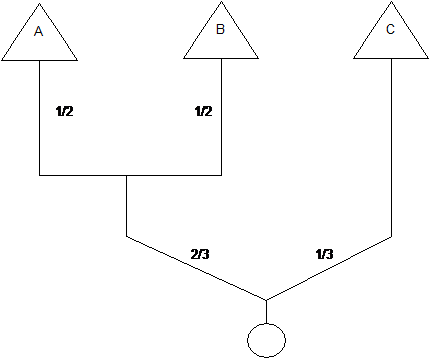}

\subsubsection{Analysis}
\begin{itemize}
\item If all players are honest, then $A$ (same for $B$) has at least\\
 $\frac{1}{2}\cdot q^{\prime}\geq\frac{1}{2}\cdot\left(\frac{2}{3}-\frac{\epsilon}{2}\right)=\frac{1}{3}-\frac{\epsilon}{4}>\frac{1}{3}-\epsilon$
chance of winning. ($\frac{1}{2}$ the chance of winning against $B$,
and $q^{\prime}$ to then win against $C$). \\
$C$ has just one game, so he obviously has at least $1-q^{\prime}\geq\frac{1}{3}-\frac{\epsilon}{2}>\frac{1}{3}-\epsilon$
chance of winning.
\item If $A$ is honest, we can think of it as if $A$ is the only honest
player. Then the calculation is almost the same from her point of
view: In the first game she has $\frac{1}{2}-\epsilon^{\prime}$ winning
probability, and in the second (conditioned that she had won the first
round) she has at least \\
$q^{\prime}-\epsilon^{\prime}\geq\frac{2}{3}-\epsilon^{\prime}-\epsilon^{\prime}=\frac{2}{3}-\epsilon$
winning probability. So in the total she has $(\frac{1}{2}-\frac{\epsilon}{2})(\frac{2}{3}-\epsilon)\geq\frac{1}{3}-\frac{5}{6}\epsilon+\frac{\epsilon^{2}}{2}>\frac{1}{3}-\frac{5}{6}\epsilon>\frac{1}{3}-\epsilon$.
\item If $B$ is honest then the calculation is the same, just replace $A$
with $B$ and vice versa.
\item If $C$ is honest - again he has only one flip, so he has at least
\\
$1-q^{\prime}-\epsilon^{\prime}\geq1-\frac{2}{3}-\epsilon^{\prime}-\epsilon^{\prime}=\frac{1}{3}-\epsilon$
chance of winning.
\item Number of coin flips = $2$.
\item First coin flip is $P_{\epsilon^{\prime}}$, hence involves $N_{\frac{\epsilon}{2}}$
rounds.\\
Second coin flip is $P_{\frac{2}{3}^{\prime},\epsilon^{\prime}}$,
and by corollary \ref{cor:P_q,e}, it involves $O\left(N_{\frac{\epsilon}{4}}\log\frac{1}{\epsilon}\right)$
rounds, hence the total running time is $O\left(N_{\frac{\epsilon}{4}}\log\frac{1}{\epsilon}\right)$.
\end{itemize}

\subsection{\label{first protocol}Simple solution}

In the general case we have $n$ parties $A_{1},\ldots,A_{n}$.

Let $\epsilon>0$. We will show a leader election protocol, such that
an honest player has a winning probability of at least $\frac{1}{n}-\epsilon$.

Let $\epsilon^{\prime}=\frac{\epsilon}{2n}$.

\subsubsection{Protocol:}
\begin{enumerate}
\item Let $W_{1}=A_{1}$.
\item For $i=2\ to\ n$

\begin{enumerate}
\item $W_{i-1}$ plays $A_{i}$ a $P_{\frac{i-1}{i}^{\prime},\epsilon^{\prime}}$
unbalanced weak coin flipping protocol. Namely, $P_{q^{\prime},\epsilon^{\prime}}$
such that $\left|q^{\prime}-\left(\frac{i-1}{i}\right)\right|\le\epsilon^{\prime}$.\\
Except when $i=2$, then the protocol is simply $P_{\epsilon^{\prime}}$.
\item $W_{i}$ is the winner.
\end{enumerate}
\item $W_{n}$ is declared as the leader.
\end{enumerate}
\includegraphics[scale=0.6]{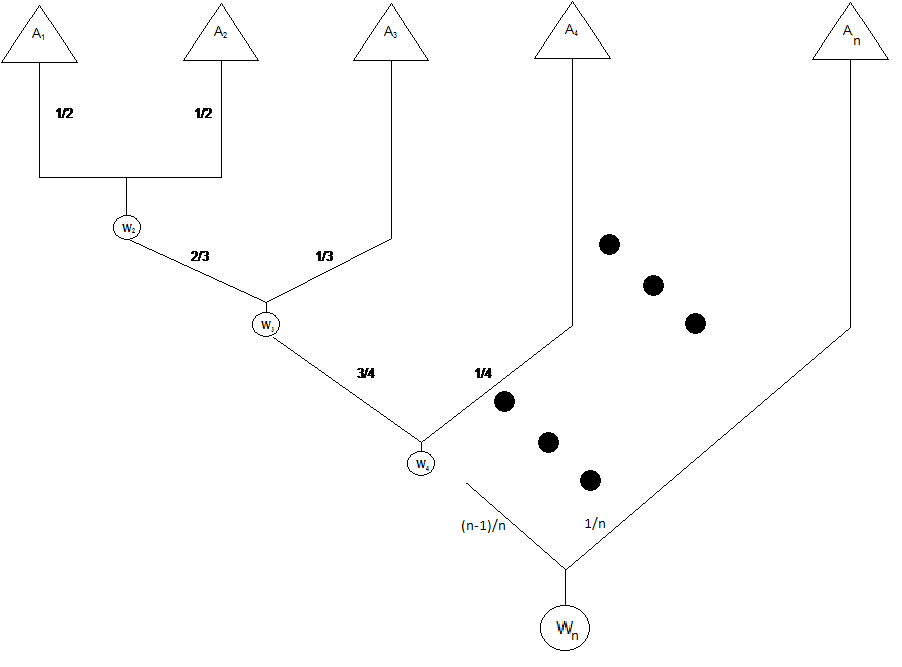}

\subsubsection{Analysis}

Note that player $j$ enters the game in the $j^{th}$ stage (i.e.
when $i=j$ on $\#2_{a}$ in the protocol), when he plays coin flip
with winning probability at least $1-q^{\prime}-\epsilon^{\prime}\geq1-\frac{i-1}{i}-2\epsilon^{\prime}\geq\frac{1}{i}-2\epsilon^{\prime}$
(The only exception in the protocol is $A_{1}$ that also plays for
the first time when $i=2$, but then also $P_{1}=$$\frac{1}{2}$,
so it has the same path as $A_{2}$).
\begin{lem}
\label{lem:(a-x)}If $\forall i\ 0\leq a_{i}\leq1,\quad0\leq x\leq\frac{1}{n}\prod_{i=1}^{n}a_{i}$
, then 
\[
\prod_{i=1}^{n}\left(a_{i}-x\right)\geq\left(\prod_{i=1}^{n}a_{i}\right)-nx
\]
\end{lem}
\begin{proof}
By induction on $n$.\\
If $n=2$ then $\left(a-x\right)\left(b-x\right)=ab-x\left(a+b\right)+x^{2}\geq ab-2x$.\\
Assume correctness for $n$ and prove for $n+1$:\\
$\prod_{i=1}^{n+1}\left(a_{i}-x\right)=\left(\prod_{i=1}^{n}\left(a_{i}-x\right)\right)\left(a_{n+1}-x\right)\geq\left(\left(\prod_{i=1}^{n}a_{i}\right)-nx\right)\left(a_{n+1}-x\right)=$\\
because every multiplicand is non-negative (from our assumptions on
$x$)\\
$=\left(\prod_{i=1}^{n+1}a_{i}\right)-x\left(n\cdot a_{n+1}+\prod_{i=1}^{n}a_{i}\right)+nx^{2}\geq\left(\prod_{i=1}^{n+1}a_{i}\right)-x\left(n+1\right)$
\end{proof}
\begin{itemize}
\item If $A_{i}$ is honest $\left(i\geq2\right)$, then his winning probability
is at least:\\
$\left(\frac{1}{i}-2\epsilon^{\prime}\right)\cdot\left(\frac{i}{i+1}-2\epsilon^{\prime}\right)\cdot\left(\frac{i+1}{i+2}-2\epsilon^{\prime}\right)\cdot\ldots\cdot\left(\frac{n-2}{n-1}-2\epsilon^{\prime}\right)\cdot\left(\frac{n-1}{n}-2\epsilon^{\prime}\right)$\\
$\geq\frac{1}{n}-\left(n-i+1\right)2\epsilon^{\prime}\geq\frac{1}{n}-\epsilon$
by lemma \ref{lem:(a-x)}, with \\
$x=2\epsilon^{\prime}=\frac{\epsilon}{n},\,a_{1}=\frac{1}{i},\,a_{j}=\frac{i-2+j}{i-1+j}\ j=2,\ldots,n-i+1$
and indeed\\
 $0<\left\{ a_{j}\right\} $ and $0<x<\prod_{j=1}^{n-i+1}a_{j}=\frac{1}{n}$.
\item If all are honest, then $A_{i}$ winning probability is only bigger
than that (we have $-\epsilon^{\prime}$ instead of $-2\epsilon^{\prime}$
at every multiplicand). 
\item Number of coin flips = $n-1$.\\
The $i^{th}$ coin flip is $P_{\frac{i-1}{i}^{\prime},\epsilon^{\prime}}$,
hence by corollary \ref{cor:P_q,e}, it will have $O\left(\log(\frac{n}{\epsilon})\cdot N_{\frac{\epsilon}{4n}}\right)$
rounds. 
\end{itemize}
This proves theorem \ref{thm2:exist}.

\section{\label{sec:Improved-protocol}Improved protocol}

We mentioned in the beginning that if $n=2^{m}$, we can do a tournament
with $m$ rounds. In the last try we came up with a protocol of $n-1$
rounds, because each time only one couple played a coin flip. The
problem with this simple solution is that it is quite inefficient
in terms of number of rounds, and also almost all the coin flips are
unbalanced, which implies an extra factor of at least $\log\frac{1}{\epsilon}$
per unbalanced coin flip to the running time. We can improve this
protocol by combining it with the tournament idea.

\subsection{Improved protocol for seven players}

We will first start by creating an efficient protocol for seven players,
before describing the general solution.

\subsubsection{Protocol}
\begin{enumerate}
\item The following couples play $P_{\epsilon}$ (balanced coin flip): \\
$A_{1}-A_{2},\ A_{3}-A_{4},\ A_{5}-A_{6}$.
\item The winners of $A_{1}-A_{2},\ A_{3}-A_{4}$ play between them $P_{\epsilon}$.
\\
The winner of $A_{5}-A_{6}$ plays $A_{7}$ a $P_{\frac{2}{3}^{\prime},\frac{\epsilon}{2}}$.
Namely, $P_{q^{\prime},\frac{\epsilon}{2}}$ such that $\left|q^{\prime}-\frac{2}{3}\right|\le\frac{\epsilon}{2}$.
\item The two winners of last stage play a $P_{\frac{4}{7}^{\prime},\frac{\epsilon}{2}}$.
\end{enumerate}
\includegraphics[scale=0.75]{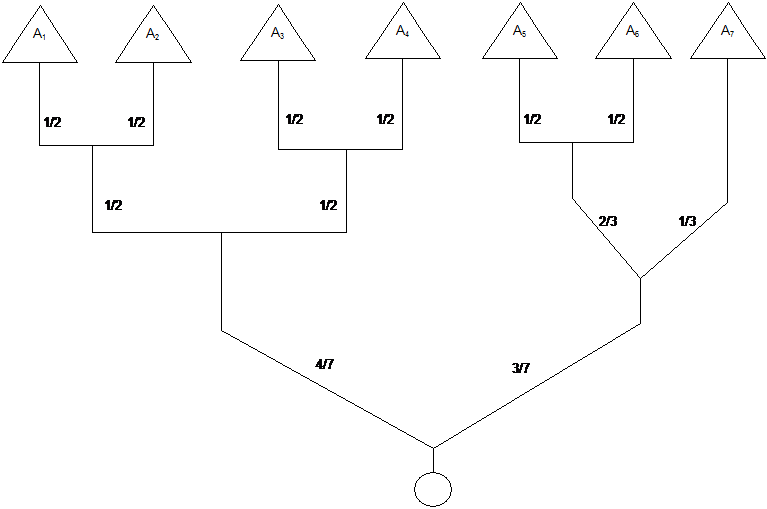}

\subsubsection{Analysis}
\begin{itemize}
\item If all are honest, then $A_{1}-A_{4}$ have the same steps, and they
have winning probability (using similar calculations as before) of
at least \\
$\frac{1}{2}\cdot\frac{1}{2}\cdot\left(\frac{4}{7}-\frac{\epsilon}{2}\right)\geq\frac{1}{7}-\frac{1}{8}\epsilon\geq\frac{1}{7}-\epsilon$.\\
$A_{5},A_{6}$ have a winning probability of at least \\
$\frac{1}{2}\cdot\left(\frac{2}{3}-\frac{\epsilon}{2}\right)\cdot\left(\frac{3}{7}-\frac{\epsilon}{2}\right)\geq\frac{1}{7}-\frac{1}{2}\epsilon\left(\frac{3}{14}+\frac{1}{3}\right)+\frac{1}{8}\epsilon^{2}\geq\frac{1}{7}-\epsilon$.\\
$A_{7}$ has only two flips, so obviously $P_{7}\geq\left(\frac{1}{3}-\frac{\epsilon}{2}\right)\cdot\left(\frac{3}{7}-\frac{\epsilon}{2}\right)\geq\frac{1}{7}-\epsilon$.
\item If $A_{1}$ is honest, he has winning probability of at least\\
$(\frac{1}{2}-\epsilon)(\frac{1}{2}-\epsilon)(\frac{4}{7}-\epsilon)=\frac{1}{7}-\frac{23}{28}\epsilon+1\frac{4}{7}\epsilon^{2}-\epsilon^{3}\geq\frac{1}{7}-\epsilon$
(since $0<\epsilon<1)$.\\
Same result for $A_{2}-A_{4}$.\\
If $A_{5}$ (or $A_{6}$) is honest then he has winning probability
of at least\\
 $(\frac{1}{2}-\epsilon)(\frac{2}{3}-\epsilon)(\frac{3}{7}-\epsilon)\geq\frac{1}{7}-\frac{35}{42}\epsilon+1\frac{25}{42}\epsilon^{2}-\epsilon^{3}\geq\frac{1}{7}-\epsilon$.\\
For $A_{7}$ it is obviously at least$\left(\frac{1}{3}-\epsilon\right)\left(\frac{3}{7}-\epsilon\right)\geq\frac{1}{7}-\frac{16}{21}\epsilon\geq\frac{1}{7}-\epsilon$.
\item Number of coin flipping rounds $=3$.
\item The two unbalanced coin flips will have each $O\left(N_{\frac{\epsilon}{4}}\cdot\log\frac{1}{\epsilon}\right)$
rounds according to corollary \ref{cor:P_q,e}, so the total protocol
has running time of $O\left(N_{\frac{\epsilon}{4}}\cdot\log\frac{1}{\epsilon}\right)$.
\end{itemize}

\subsection{Final protocol\label{subsec:Final-protocol}}

In the general case we have $n$ parties $A_{1},\ldots,A_{n}$. \\
Let $\epsilon>0$. We will show a leader election protocol $LE_{\epsilon}$,
with $\log(n)$ coin flipping rounds, and a running time of $O\left(N_{\frac{\epsilon}{4}}\cdot\log n\cdot\log\frac{1}{\epsilon}\right)$,
s.t. $\forall i$ if party $i$ is honest, he has winning probability
of at least $\frac{1}{n}-\epsilon$.

\subsubsection{Protocol}

We will define the protocol recursively.

Let us call it $Leader\left(\{A_{1},\ldots,A_{n}\},\epsilon\right)$. 

Say it returns the leader selected.

Let $k\in\mathbb{N}$ s.t. $2^{k}\leq n<2^{k+1}$.
\begin{enumerate}
\item The following are done simultaneously:

\begin{itemize}
\item $A_{1},\ldots,A_{2^{k}}$ plays a tournament (of $k$ rounds) among
themselves with $P_{\frac{\epsilon}{2}}$. Denote the winner as $w_{1}$.
\item $w_{2}=Leader\left(\{A_{2^{k}+1},\ldots,A_{n}\},\epsilon\right)$.
\end{itemize}
\item $w_{1}$ plays $w_{2}$ a $P_{\frac{2^{k}}{n}^{\prime},\frac{\epsilon}{2}}$.
Namely, $P_{q^{\prime},\frac{\epsilon}{2}}$ such that $\left|q^{\prime}-\frac{2^{k}}{n}\right|\le\frac{\epsilon}{2}$.\\
\\
The winner of this is the leader.
\end{enumerate}
\begin{rem*}
In contrast to the protocol presented with seven players, here the
$2^{k}$ first players play $P_{\frac{\epsilon}{2}}$ between themselves.
If we had used weak coin flipping protocol $P_{\epsilon}$, this would
give a Leader Election protocol for $n$ players with bias a bit over
$\epsilon$ for $n=3,5$. However, this small concession doesn't increase
our running time.
\end{rem*}

\subsubsection{Analysis}
\begin{itemize}
\item Assume that all parties are honest. We shall prove by induction the
following:
\begin{claim}
If all parties are honest, then $P_{i}\geq\frac{1}{n}-\epsilon$.
\end{claim}
\begin{proof}
For $n=2$ it is obvious. Assume correctness for all $m<n$, and we
will prove it for $n$.\\
Look at $A_{1},\ldots,A_{2^{k}}$. In the tournament everyone has
a $\frac{1}{2^{k}}$ winning chance. Whoever wins, has another coin
flip with at least $\left(\frac{2^{k}}{n}-\frac{\epsilon}{2}\right)$
winning chance, so in total they each have at least\\
$\frac{1}{2^{k}}\cdot\left(\frac{2^{k}}{n}-\frac{\epsilon}{2}\right)=\frac{1}{n}-\frac{\epsilon}{2^{k+1}}\geq\frac{1}{n}-\epsilon$
winning chance.\\
From the induction hypothesis, we know that each one of $A_{2^{k}+1},\ldots,A_{n}$
has at least $\frac{1}{n-2^{k}}-\epsilon$ winning chance in the recursive
leader election procedure (to become $w_{2}$). Notice that $\frac{1}{n-2^{k}}$
is well defined since we can assume that $n$ is strictly larger than
$2^{k}$, otherwise we are done. Then the winner has at least $\left(1-\frac{2^{k}}{n}-\frac{\epsilon}{2}\right)$
winning chance in the last step, so altogether he has a winning probability
$P_{i}\geq\left(\frac{1}{n-2^{k}}-\epsilon\right)\cdot(1-\frac{2^{k}}{n}-\frac{\epsilon}{2})=\frac{1}{n}-\epsilon\left(\frac{1}{2}\frac{1}{n-2^{k}}+1-\frac{2^{k}}{n}\right)+\frac{\epsilon^{2}}{2}$\\
since $2^{k}>\frac{n}{2}$ we get that $P_{i}\geq\frac{1}{n}-\epsilon$$\left(\frac{1}{2}+1-\frac{1}{2}\right)$\\
hence $P_{i}\geq\frac{1}{n}-\epsilon$.
\end{proof}
\item We have $\log(n)$ coin flipping rounds, and according to corollary
\eqref{cor:P_q,e} we will have up to $O(N_{\frac{\epsilon}{4}}\log(\frac{1}{\epsilon})\log(n))$
total rounds.
\item The number of unbalanced coin flips is bounded by $log(n)$.
\begin{claim}
$\#$unbalanced coin flips $=$ (\# of $1$'s in the binary representation
of $n$) - 1.
\end{claim}
\begin{proof}
This can be proved easily by induction on $n$:\\
For $n=1,2$ it is clear. No unbalanced coin flipping protocols are
used. \\
If $n$ is a power of $2$, say $n=2^{k}$, then it has $0$ such.
\\
Else $2^{k}<n<2^{k+1}$, and the first $2^{k}$ players use again
$0$ unbalanced coin flips between them. The remaining $m=n-2^{k}$
players use (from the induction hypothesis) the $\#$ of $1$'s in
the binary representation of $m$, minus $1$, in the appropriate
rounds. When joining the two groups, we again use an unbalanced coin
flip (in the last round) which corresponds to the MSB of $n$. As
$n=2^{k}+m$, $\#1$'s in $m$ is exactly one less than the $\#$
of $1$s in $n$, which proves the claim. 
\end{proof}
\begin{rem*}
There can only be one unbalanced coin flip per round $i$, if the
$i^{th}$ bit in the binary representation of $n$ is $1$.
\end{rem*}
\item The last thing needed to be proven is that an honest player has a
winning probability of $\frac{1}{n}-\epsilon$.\\
A simple proof of that will be: \\
If $A_{i}$ is honest, then he has a winning probability of at least\\
$\underset{i=1}{\overset{\log(n)}{\prod}}(c_{i}-\epsilon)\geq\frac{1}{n}-\log(n)\epsilon$
where the $\left\{ c_{i}\right\} $ are the winning probabilities
of $A_{i}$ in the individual weak coin flipping rounds ( $c_{i}<1$
)\\
Therefore, we can use $\frac{\epsilon}{\log n}$ as the weak unbalanced
coin flipping bias in the protocol (instead of $\frac{\epsilon}{2}$).\\
Then we finish the proof, but we get running time of $O\left(N_{\frac{\epsilon}{2\log n}}\log n\log\frac{\log n}{\epsilon}\right)$
instead of $O\left(N_{\frac{\epsilon}{4}}\log n\log\frac{1}{\epsilon}\right)$.\\
While one might say that $\log\log n$ factor is insignificant, the
difference between $N_{\frac{\epsilon}{4}}$ and $N_{\frac{\epsilon}{2\log n}}$
might be huge.\\
In fact, at the moment, the only known proof for existence of weak
coin flipping with arbitrarily small bias $P_{\epsilon}$ is analyzed
in \cite{DoritAharonov2016}, giving a bound of $N_{\epsilon}\leq\frac{1}{\epsilon}^{O\left(\frac{1}{\epsilon}\right)}$
rounds.

Therefore, we will make a more precise calculation, which will allow
us to use our original $\frac{\epsilon}{2}$ in the unbalanced coin
flip.
\item Let us first do a precise calculation for $n=3$:\\
$A_{1}$ will play $P_{\frac{\epsilon}{2}}$ with $A_{2}$, and the
winner will play $P_{\frac{2}{3}^{\prime},\frac{\epsilon}{2}}$.\\
An honest $A_{1}$ (or $A_{2}$) will have a winning probability of
\\
$\left(\frac{1}{2}-\frac{\epsilon}{2}\right)\left(\frac{2}{3}-\epsilon\right)\geq\frac{1}{3}-\frac{5}{6}\epsilon\geq\frac{1}{3}-\epsilon$.\\
The running time of this protocol is $O\left(N_{\frac{\epsilon}{4}}\cdot\log\frac{1}{\epsilon}\right)$,
by corollary \eqref{cor:P_q,e}.
\end{itemize}
We now prove for n players: 
\begin{lem}
An honest player $A_{i}$ has a winning probability which is at least
$\frac{1}{n}-\epsilon$.
\begin{proof}
We will prove it by induction on $n$.\\
If $n=1,2,3,4$ then we saw this is true.\\
Assume correctness for all $n<N$, and we shall prove it for $n=N$.
\begin{itemize}
\item We saw this is true if $N$ is a power of $2$.
\item If $N=2^{k}+1$:\\
Assume $A_{N}$ is honest. $A_{N}$ only plays one unbalanced coin
flip $P_{\frac{2^{k}}{N}^{\prime},\frac{\epsilon}{2}}$, having winning
probability of at least $\frac{1}{N}-\epsilon$ by corollary \eqref{cor:P_q,e}.\\
$A_{1},\ldots,A_{N-1}$ play a full tournament of $2^{k}$ players,
with winning probability of $p\geq\left(\frac{1}{2}-\frac{\epsilon}{2}\right)^{k}=\left(\frac{1-\epsilon}{2}\right)^{k}=\frac{\left(1-\epsilon\right)^{k}}{N-1}$
for an honest player. \\
By Bernoulli inequality:$\forall x\geq-1,n\in\mathbb{N}\ \left(1+x\right)^{n}\geq1+nx$,
plugging $x=-\epsilon$, we get that $p\geq\frac{1-\epsilon k}{N-1}$.\\
Then the honest winner plays $P_{\frac{2^{k}}{N}^{\prime},\frac{\epsilon}{2}}$
with $A_{N}$, hence having winning probability of at least 
\[
\frac{1-\epsilon k}{N-1}\left(\frac{N-1}{N}-\epsilon\right)=\frac{1}{N}-\epsilon\left(\frac{k}{N}+\frac{1}{N-1}\right)+\epsilon^{2}\frac{NK}{N-1}\geq\frac{1}{N}-\epsilon
\]
Where the last inequality is true for every $N=2^{k}+1\geq3$ (because
$k=\left\lfloor \log N\right\rfloor $, and already for $k=1$ we
get $\frac{k}{N}+\frac{1}{N-1}=\frac{1}{3}+\frac{1}{2}<1$).\\
\item Otherwise $2^{k}+2\leq N<2^{k+1}$, for $k=\left\lfloor \log N\right\rfloor \,,\ k\ge2$.\\
For all $i\leq2^{k}$ by the induction hypothesis (for $n=2^{k}$),
we have that the probability for $A_{i}$ to reach the final weak
coin flipping round (round $k+1$) is at least $\frac{1}{2^{k}}-\epsilon$.\\
Hence, his winning probability (to be elected) $p$ is at least 
\[
p\geq\left(\frac{1}{2^{k}}-\epsilon\right)\left(\frac{2^{k}}{N}-\epsilon\right)=\frac{1}{N}-\epsilon\left(\frac{2^{k}}{N}+\frac{1}{2^{k}}\right)+\epsilon^{2}
\]
Since $2^{k}+2\leq N<2\cdot2^{k}$ we get that $p>\frac{1}{N}-\epsilon$$\left(\frac{2^{k}+2}{N}\right)\geq\frac{1}{N}-\epsilon$.\\
For all $i>2^{k}$ by the induction hypothesis (for $t=N-2^{k}$),
we have that the probability for $A_{i}$ to reach the final weak
coin flipping round is at least $\frac{1}{t}-\epsilon$.\\
Hence, his winning probability (to be elected) is at least 
\[
\left(\frac{1}{t}-\epsilon\right)\left(\frac{t}{N}-\epsilon\right)\geq\frac{1}{N}-\epsilon\left(\frac{t}{N}+\frac{1}{t}\right)+\epsilon^{2}\geq\frac{1}{N}-\epsilon\left(\frac{t}{N}+\frac{1}{t}\right)
\]
Since $\frac{t}{N}\leq\frac{1}{2}\,,\ t\geq2$ we get that $p\geq\frac{1}{N}-\epsilon$.
\end{itemize}
\end{proof}
\end{lem}
This completes the proof of theorem \ref{thm3:efficient}.

\section{Open questions\label{sec:Open-questions}}

Here we will present the open question rising from this paper.
\begin{itemize}
\item The first obvious open question, is to find a specific family of balanced
weak coin flipping $P_{\epsilon}$, hence getting a tighter bound
on $N_{\epsilon}$.
\item The second open question, which was mentioned before, is to try and
find unbalanced weak coin flipping directly, hence removing the $O\left(\log\frac{1}{\epsilon}\right)$
factor from corollary \ref{cor:P_q,e}.
\item A more specific question is: Can one improve the three party leader
election protocol. Our solution (subsection \ref{subsec:Final-protocol})
used running time of $O\left(N_{\frac{\epsilon}{4}}\cdot\log\frac{1}{\epsilon}\right)$,
and the number of rounds in $P_{\frac{\epsilon}{4}}$ might be significantly
greater than in $P_{\epsilon}$.
\item Can one find a better solution (in terms of running time) to the leader
election problem? Maybe by finding a direct solution to the problem,
and not via weak coin flipping.
\item When we needed to use unbalanced weak coin flipping $P_{q,\epsilon}$,
we used the approximate coin flip $P_{q^{\prime},\frac{\epsilon}{2}}$,
which implied two annoying consequences:
\begin{itemize}
\item If all players are honest, they are not granted a $\frac{1}{n}$ winning
probability, but only $\frac{1}{n}-\frac{\epsilon}{2}$.\\
Hence the open question is, can one find an arbitrarily small biased
leader election, that guarantees exactly $\frac{1}{n}$ winning probability
in the scenario where all players are honest?
\item Can one improve corollary \ref{cor:P_q,e} to use $O\left(N_{\delta}\cdot\log\frac{1}{\delta}\right)$
rounds, which will improve Theorem \ref{thm3:efficient} running time
to $O\left(N_{\frac{\epsilon}{2}}\cdot\log n\cdot\log\frac{1}{\epsilon}\right)$?
\end{itemize}
\end{itemize}

\section*{$\qquad\qquad\qquad$Acknowledgments}

I would like to thank my adviser Prof. Dorit Aharonov for her support
and guidance, and for her remarks on this paper (in its many versions),\\
and also thank Prof. Michael Ben-Or for his help.

\appendix

\section*{\label{sec:Remarks}Appendix: Weak coin flipping}

Let $P$ be a weak coin flipping protocol, with $P_{B}^{*}$ the maximal
cheating probability of Bob.

We want to run two instances of $P$, one after the other (not even
at the same time).

We will define (see \cite{mochon-2007,DoritAharonov2016} for full
details) for $P$:
\begin{itemize}
\item Let $\mathcal{H=A\otimes M\otimes B}$ be the Hilbert space of the
system.
\item $\mid\psi_{0}>=\mid\psi_{A,0}>\mid\psi_{M,0}>\mid\psi_{B,0}>$ is
the initial state of the system.
\item Let there be $n$ (even) stages, $i$ denote the current stage.
\item On the odd stages $i$, Alice will apply a unitary $U_{A,i}$ on $\mathcal{A\otimes M}$.
\item On the even stages, Bob will apply a unitary $U_{B,i}$ on $\mathcal{M\otimes B}$.
\item Let $\mid\psi_{i}>$ be the state of the system in the $i_{th}$ stage.
\item Let $\rho_{A,i}=Tr_{\mathcal{M\otimes B}}(\mid\psi_{i}><\psi_{i}\mid)$
be the density matrix of Alice in the $i_{th}$ stage.
\item Alice's initial state (density matrix) is $^{(i)}\rho_{A,0}=|\psi_{A,0}><\psi_{A,0}|$.
\item For even state $i$ we have $^{(ii)}\rho_{A,i}=\rho_{A,i-1}$.
\item Let $\tilde{\rho}_{A,i}$ be the state of $\mathcal{A\otimes M}$
after Alice gets the $i_{th}$ message.
\item For odd $i$ : $^{(iii)}\rho_{A,i}=Tr_{\mathcal{M}}(\tilde{\rho}_{A,i})$,
$^{(iv)}\rho_{A,i}=Tr_{\mathcal{M}}(U_{A,i}\tilde{\rho}_{A,i-1}U_{A,i}^{\dagger})$.
\end{itemize}
We know that regardless of Bob's actions (see \cite{mochon-2007,DoritAharonov2016}
for full proof):
\begin{equation}
P_{B}^{*}\leq\max Tr[\Pi_{A,1}\rho_{A,n}]\label{Bob's bound}
\end{equation}
where the maximization is done over all density matrices $\rho$ that
satisfies the conditions $\left(i\right)-\left(iv\right)$.

\bibliographystyle{amsalpha}
\bibliography{Leader_election}

\end{document}